# Local breaking of the spin-orbit interaction: the microscopic origin of exchange bias in Co/FeMn

Sebastian Brück[1,2], Patrick Audehm[1], Gisela Schütz[1], and Eberhard Goering[1]

[1]*Max-Planck-Institut für Metallforschung, Heisenbergstrasse 3, D-70569 Stuttgart*

[2]*Physikalisches Institut der Universität Würzburg, Am Hubland, D-97074 Würzburg*

Abstract:

Modern magnetic thin film devices owe their success in large part to effects emerging from interlayer coupling and exchange interaction at interfaces. A prominent example is exchange bias (EB), a magnetic coupling phenomenon found in ferromagnet (F)/antiferromagnet (AF) systems. Uncompensated pinned moments in the AF couple to the F via the interface causing an additional unidirectional anisotropy[1,2]. As a result, the hysteresis of the F is shifted. The existence of such pinned moments is nowadays accepted although their physical nature and origin is still unknown. Here we present a thorough spectroscopic investigation based on X-ray magnetic circular dichroism which does for the first time provide direct information about the physics of pinned magnetic moments. Our data clearly shows that the orbital magnetic moment, which is usually widely quenched in transition metal systems, is the driving force behind exchange bias in Co/FeMn.

During the last years the EB discussion was focused on the role of the interface disorder and roughness for the coupling of a pinned moment in the AF with the F magnetic spin moment[3,4]. Most of these considerations assumed that the involved magnetic moment in the AF also originates from the spin. This assumption was driven by intuition since the orbital magnetic moment is usually quenched in transition metal systems. However, an isotropic (spin) moment cannot cause any unidirectional symmetry breaking effect like EB. Orbital magnetic moments on the other hand may exhibit a strong anisotropic behaviour and their role for EB should be fundamental. Hence a closer examination of the role of the two species of magnetic moments for EB is indispensable. We use X-ray magnetic circular dichroism (XMCD) combined with the interface sensitive X-ray resonant magnetic scattering (XRMS) to identify the role of orbital magnetism for the exchange bias effect.

Two polycrystalline Co/FeMn samples with varying FeMn thickness were investigated by these techniques. The thickness of the antiferromagnet is varied to modify the blocking temperature in the system and by this the onset of the exchange bias loop shift[5,6]. Sample #1 (50Å FeMn) reveals no EB at RT as can be seen from the SQUID hysteresis shown as an inset in Figure 2c). However, after field cooling the sample to T=50K in a negative external field of -65mT, the hysteresis loop (blue) exhibits a large exchange bias of $H_{eb}$=+29.3mT antiparallel to the cooling field. The second sample, #2 (100Å FeMn) shows room temperature EB and the SQUID hysteresis (blue) shown as an inset in Figure 3 exhibits a moderate exchange bias of $H_{eb}$=+4.5mT at RT.

X-ray absorption and energy dependent scattering at a fixed momentum transfer $q_z$ were measured at the UE56/2-PGM1 beamline at BESSY II in Berlin using a dedicated UHV absorption and reflectometry end station[7]. By adjusting the energy of the incident monochromatic X-rays to the magnetic active 2p → 3d transition (L edges) of Fe it is possible to characterize the signal of uncompensated rotatable and pinned Fe moments in the antiferromagnetic FeMn separately.

Energy dependent resonant magnetic scattering at the Fe L edges of sample #1 was performed at a constant $q_z$ value of $q_z=0.155\text{Å}^{-1}$ for flipped magnetization and/or flipped circular polarization of the X-rays. The resulting 4 RT curves according to the schemes ↑↑, ↑↓, ↓↑, and ↓↓ are shown in Figure 2a). Here, the first arrow indicates the helicity of the X-rays, positive ↑ or negative ↓, while the second one is related to the magnetic field direction. The two resonant absorption lines $L_2$ and $L_3$ are clearly visible in the spectra although the general shape differs significantly from an absorption spectrum due to the mixture of dispersion and absorption in reflection measurements. From these four curves, the signal of both species of uncompensated moments in the antiferromagnet is extracted according to the schemes illustrated in Figure 1.

Rotatable Fe moments are obtained by applying an XMCD like calculus of the form: $I_{rotatable}$ = ↑↑-↑↓ (see also Figure 1). Figure 2b) shows the resulting XRMR signal $I_{rotatable}$ of rotatable uncompensated Fe in the antiferromagnet at 300K and after field cooling the sample to 50K.

A large ferromagnetic signal indicating the presence of rotatable uncompensated Fe moments in the sample is found. The existence of rotatable moments at the ferromagnet/antiferromagnet interface has already been proven for a variety of exchange bias samples including FeMn[3,8,9]. The general shape of both curves resembles a well known Fe metal like XMCD signal. Especially the sign change between $L_3$ and $L_2$ edges is typical for a spin moment dominated XMCD[10].

It is in general not possible to identify a signal from pinned uncompensated moments by flipping the external field, because of their pinned nature as illustrated in Figure 1. To detect pinned moments the following alternative calculus has to be used: $I_{pinned}$ = ↑↑-↓↓ (or: $I_{pinned}$ = ↑↓-↓↑). This calculus ensures that the rotatable moments are always kept parallel (or antiparallel) with respect to the helicity of the incident X-rays and thus effectively blanks out their contribution (see Figure 1). The remaining $I_{pinned}$ XMCD signal is caused by pinned uncompensated moments, which change from parallel to antiparallel by the change in the light helicity. Applying this scheme to the 300K resonant scattering curves from Figure2a) and their 50K counterparts leads to the two curves shown in Figure 2c). The signal originated from pinned uncompensated Fe moments in the antiferromagnet is virtually zero at room temperature (red line). But at 50K, thus after inducing EB, a pronounced signal is found. This clearly indicates a direct link between this signal and the exchange bias loop shift.

By comparing the shape of the pinned difference signal with the corresponding signal from rotatable Fe a fundamental difference is evident: While the latter shows a sign change between $L_3$ and $L_2$, the pinned difference signal exhibits the same sign for $L_3$ and $L_2$.

The spectral line shape of the 2p → 3d absorption and by this also the shape of a difference signal like $I_{rotatable}$ or $I_{pinned}$ is based on ground state expectation values, like orbital and spin moments. Spin moments reveal opposite sign with equal area at the $L_2$ and $L_3$ edges, while orbital moments reveal the same sign with double integral intensity at the $L_3$ edge with respect to the $L_2$ edge[11]. This is the basis for the well known XMCD sum rules, which allow the seperation of the spin and orbital magnetic moments in XMCD spectra[10,12]. Therefore the general shape of the pinnend moment difference signal with no sign change clearly suggests a dominating contribution of pinned orbital moments, while the rotatable signal is spin dominated[11].

To verify the observed line shape from Figure 2c) and perform a second test of its correlation with EB, sample #2 was investigated using the same resonant scattering technique. As already mentioned, sample #2 exhibits a room temperature exchange bias of $H_{eb}$=+4.5mT.

Instead of heating the sample beyond the AF ordering temperature, the unidirectional character of EB[1] was used to compare between no EB and $H_{eb}$=+4.5mT by measuring first parallel to the $H_{eb}$ direction and then at an azimuthal angle of 90°, where the scalar product between the circular polarization and the pinnend magnetic moments is always zero. The inset of Figure 3 shows SQUID hystereses measured parallel and perpendicular to $H_{eb}$. As expected the loop shift vanishes for the perpendicular direction and the hysteresis shows a hard axis characteristic.

The difference signal associated to pinned uncompensated Fe moments in the antiferromagnet was measured for both azimuthal angles to check the link of the EB loop shift and the orbital moment like pinned Fe spectra. Again a direct correlation of the pinned signal and $H_{eb}$ is found. The spectral line shape of the pinned difference signal exhibits the same orbital moment dominated shape, as was found for sample #1.

In a last step, the temperature behaviour of the pinned Fe moments in sample #2 was investigated by conventional XMCD spectroscopy via total electron yield (TEY). This method provides quantitative information for the spin and orbital magnetic momentum directly from the measured curves. Therefore the TEY XMCD of the Fe moments was measured for both field directions and X-ray polarizations to obtain the normal rotatable XMCD as well as the difference signal associated with the pinned Fe moments. Since the relevant Co/FeMn interface is deeply buried, the absolute amplitude of the Fe $L_{2,3}$ signal is extremely small which makes these measurements very demanding and time consuming.

Figure 4a) shows the edge normalized 2p→3d absorption edges of Fe for parallel (↑↑) and antiparallel (↑↓) alignment of the positive X-ray polarization and the applied magnetic field at room temperature. After measuring the same for the negative helicity, the sample was cooled to T=197K in the remnant state and the measurement was repeated. Note that the remnant state imprints a positive exchange bias here[13].

The corresponding difference, i.e. XMCD signal according to the rotatable moments is shown for RT and 197K in Figure 4b). A pronounced, nearly temperature independent XMCD signal is found indicating that a large portion of the Fe directly at the interface to the Co is rotatable, i.e., behaves ferromagnetic. Due to the deeply buried FeMn/Co interface, the TEY signal is dominated by Fe located close to the interface with the F. The corresponding fluorescence yield (not shown), which probes the whole Fe layer, shows a 4 times smaller XMCD signal proving that the rotatable uncompensated Fe moments are located close to the upper interface of the FeMn layer. By applying XMCD sum rules, the spin and orbital moment for the rotatable Fe in the antiferromagnet are calculated from the data in Figure 4b). From the RT measurement, an averaged spin moment of $m_S = 1.07 \mu_B$ and an orbital moment of $m_{Orb} = 0.12 \mu_B$ is found, while the T=197K data leads to a spin moment of $m_S = 1.03 \mu_B$ and an orbital magnetic moment of $m_{Orb} = 0.21 \mu_B$. The enhanced L/S ratio, compared to bulk bcc Fe[10], already indicates that the interface near region resembles reduced symmetry, which is necessary to provide enhanced Fe orbital magnetic moments. Note that these values are corrected for the angle of incidence of 60°, the beamline degree of polarization of 90% and the number of holes for Fe of $n_h = 10 - n_{3d} = 3.31$[10]. Compared to bulk values[10], the spin moment is too small. This can be explained by the above mentioned sensitivity range of the TEY: If the rotatable uncompensated Fe is confined to the interface near region, the TEY also probes a fraction of the AF Fe atoms resulting in reduced sum rules values.

Another very important result from the XMCD sum rules is the magnitude of the orbital moment compared to the spin moment. While the expected orbital to spin magnetic moment ratio for Fe should be roughly $m_{Orb}/m_S = 0.04$ according to reference data for bcc Fe[10], the ratio found here is $m_{Orb}/m_S = 0.11$ at RT and increases further to $m_{Orb}/m_S = 0.20$ when field cooling to T=197K. The relative orbital moment of the rotatable uncompensated Fe moments is strongly enhanced suggesting a locally distorted Fe environment. It is also important to note that the absolute rotatable spin moment does not change significantly during cooling. This indicates that only a nearly vanishing amount of originally rotatable uncompensated Fe spin moments is frozen in during the cooling and hence does not contribute to the pinned moment fraction. This is consistent with an absence of a sizable pinned spin moment, even at low temperatures.

In order to prove the orbital moment character, the signal from the pinned moments is also extracted from the measured TEY XAS curves. Difference signals according to the scheme (↑↑-↓↓) from the RT (red) and T=197K (blue) measurements are shown in Figure 4c). The corresponding hystereses were measured by XRMR at the Co $L_3$ edge and are shown in the inset of Figure 4c). They verify that the exchange bias is increasing during the cooling process. The signal from pinned uncompensated Fe moments is by a factor of 10 smaller than the one from the rotatable Fe. Despite the higher noise level, the general shape is clearly visible. The signal from the pinned Fe moments has the same "orbital" shape as the difference signals from Figure 2c) and Figure 3. However in contrast to the latter two, the data shown in Figure 4c) can be analyzed quantitatively by using sum rules[12,14]. A vanishing pinned spin magnetic moment of $\tilde{m}_S = (0.004 \pm 0.007)\mu_B$ and a much larger orbital moment of $\tilde{m}_{Orb} = (0.015 \pm 0.003)\mu_B$ are found from the RT curve. After field cooling, both pinned moments increase. At T=197K a pinned magnetic spin moment of $\tilde{m}_S = (0.02 \pm 0.03)\mu_B$ is found while the pinned orbital magnetic moment is $\tilde{m}_{Orb} = (0.068 \pm 0.017)\mu_B$. The realistic error bars, especially of the spin moment, are related to the presence of small "offset" signals, clearly visible in Figure 4c), which strongly affect the spin moment determination.

From the hystereses shown in the inset of Figure 4c), an increase of the EB from +4.5mT at RT to $H_{eb}$=+7.1mT at T=197K is found. At the same time the pinned total magnetic moment increases from $0.015\mu_B$ to $0.068\mu_B$.

From the resonant scattering and the absorption measurement at the Fe L edges it is obvious that the pinned magnetic Fe moment in the AF is dominated by the orbital moment and not, as usually in bulk materials, by the spin moment. The fact that the orbital moment is even larger than the spin moment provides a second, even more important conclusion: The collinearity between the expectation values for spin and orbital momentum is broken up in the pinned atoms and the spin moment is rotated away with respect to the pinned orbital moment. Remember that XMCD probes projections of moments along the beam direction which is here the EB direction. A simple reasoning leads to this unexpected finding: Even if assuming a Hund like behaviour for the Fe atoms, the Fe spin moment should always be large with respect to the orbital moment ($m_{Orb}/m_S = 1$ for 3d$^7$ or $m_{Orb}/m_S = 0.5$ for 3d$^6$). The only possible conclusion for the observed situation with much larger orbital moment is that spin and orbital momentum are effectively not collinear here. Such a behaviour was already predicted theoretically[15,16] and, for small angles, also observed experimentally in the context of the magnetocrystalline anisotropy[17]. To rotate the spin away from the orbital moment, the local Fe spin-orbit coupling energy has to be paid, resulting in a loaded intra-atomic spring. This is then the origin of the EB loop shift, i.e. the unidirectional anisotropy.

One possible explanation for these observations of orbital moments dominated pinned magnetism is based on local symmetry breaking close to the AF/F interface. This directly provides less quenched orbital magnetic moments with strongly enhanced anisotropy energy. Rotating the F moments on top of the AF layer also rotates the Fe spin moments via strong direct and/or indirect exchange coupling. The local anisotropy energy must be strong enough here to pin the orbital moments against the spin orbit interaction of the rotated spin moment. Then the tilting between spin and orbital momentum at a single Fe site stores LS-coupling energy like a loaded spring. We propose a small number of unidirectional oriented pinned orbital moments and the resulting spin-orbit coupling as the driving force for EB.

This orbital momentum dominated XMCD, while the spin is merely not visible, has not been observed so far. This idea of LS-based EB is new and in strong contrast to all exchange bias models proposed so far, where the loaded spring is related to the angle difference between the spins in the FM with respect to the spins in the AF and the force is based on exchange interaction. After more than 50 years of EB research, this observation could be solving the EB puzzle and be the basis for new quantitative models.

**Methods**

Both samples were prepared by ion-assisted RF sputtering in a custom UHV chamber with a base pressure of better than $5 \cdot 10^{-8}$ mbar. The average deposition rate under working conditions is roughly 1 Å/s and the Ar partial pressure is $1 \cdot 10^{-4}$ mbar during growth. Si pieces with a native oxide layer of roughly 2.7nm cut from a commercially bought wafer were used as substrate. A Cu buffer layer was grown on the Si to promote fcc growth of the subsequent FeMn layer and both samples were capped by an Al protection layer to prevent oxidation. The nominal layer stacking is Si/10nm Cu/5nm $Fe_{50}Mn_{50}$/6nm Co/1.5nm Al for sample #1, and Si/100nm Cu/10nm Fe50Mn50/10nm Co/2nm Al for sample #2.

The magnetometry investigation was carried out using a Quantum Design MPMS-XL 5 Tesla SQUID magnetometer. All absorption and scattering measurements were conducted during three beamtimes at the UE56/2-PGM1 beamline at BESSY II using a custom build UHV reflectometer experiment[7]. The integration time per point for the energy dependent scattering spectra was 400ms while for the absorption measurement, the integration time was 800ms per point.

**Acknowledgements**

We thank D. Steiauf and M. Fähnle for helpful discussions. Moreover we are very grateful to B. Ludescher for preparing the excellent Co/FeMn samples, and B. Zada and W. Mahler for the support during the beamtime at BESSY.

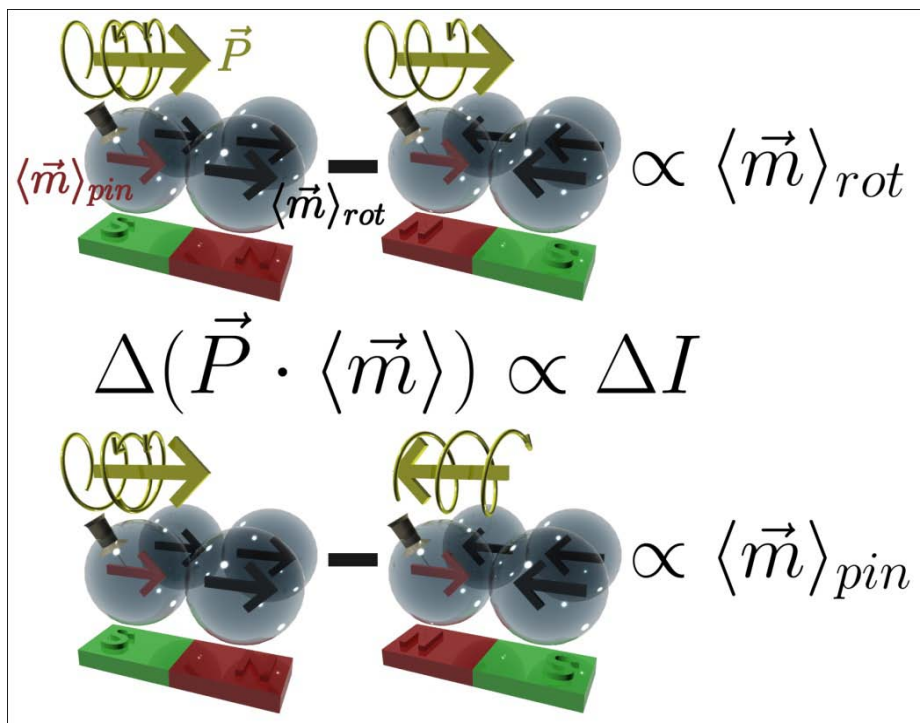

Figure 1: Illustration of the two measurement schemes. The green red magnet below symbolises the direction of the external magnetic field and the corresponding Co magnetization. The yellow arrow indicates the projected circular polarization direction of the incident X-rays. XMCD is only sensitive to changes of the scalar product between circular polarization and magnetic moment. Hence in the upper row only rotatable magnetic moments contribute to the difference signal while for the scheme described in the lower row, only pinned magnetic moments cause a difference signal.

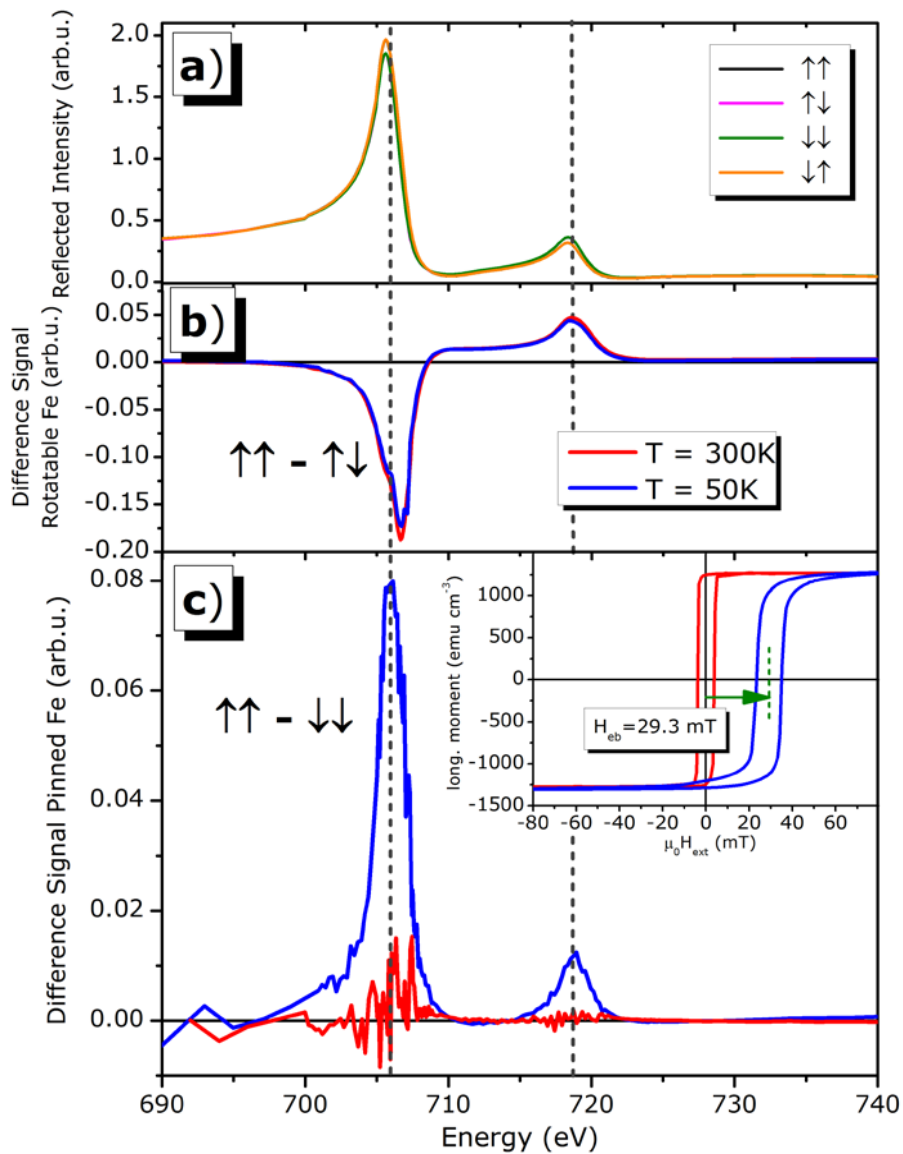

Figure 2: Temperature dependent signal from rotatable and pinned Fe moments. Resonant scattering at a constant $q_z$ of 0.155Å$^{-1}$ measured at the Fe L edges using circular polarized X-rays. The topmost graph, a), shows the four resonant scattering curves at RT measured for parallel and antiparallel alignment of the magnetization (second arrow) and the helicity of the incident X-rays (first arrow). The middle graph shows the difference signal at room temperature and at T=50K according to the scheme (↑↑-↑↓). This signal verifies that a sizable amount of uncompensated rotatable Fe moments is present in the antiferromagnet. Subfigure c) shows the signal from pinned uncompensated Fe moments

in the antiferromagnet (↑↑-↓↓) for both temperatures. The inset shows the corresponding hysteresis loops measured by SQUID magnetometry.

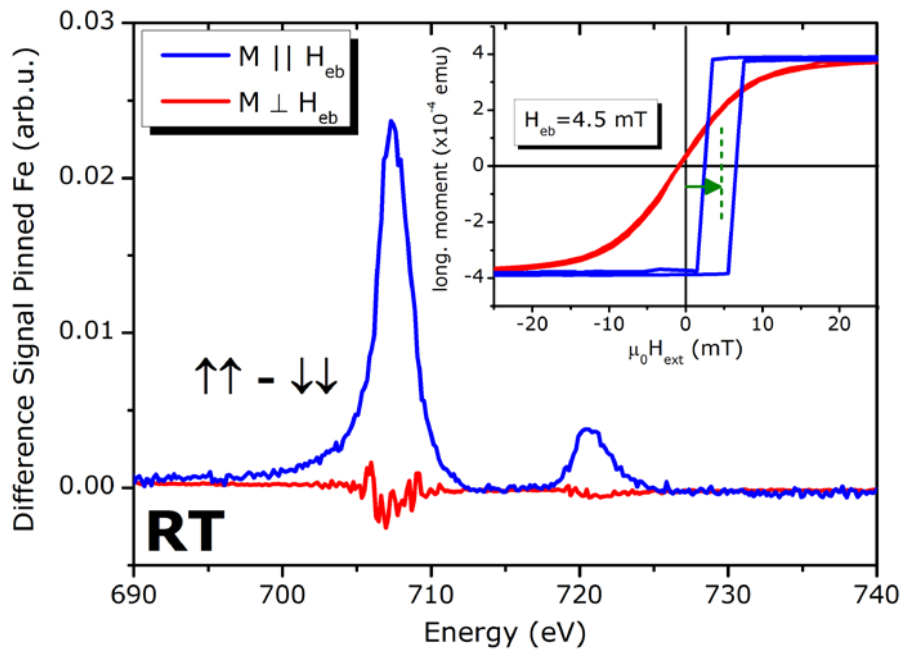

Figure 3: Azimuthal dependence of the XRMR difference signal corresponding to the pinned uncompensated Fe moments in sample #2. The red curve has been measured perpendicular to the exchange bias direction while the blue curve is measured parallel. The resonant scattering is measured at $q_z$ = 0.110Å$^{-1}$ and the difference signals were calculated according to (↑↑-↓↓). The inset shows the corresponding SQUID hysteresis loops parallel and perpendicular to $H_{eb}$. As expected from the unidirectional character of exchange bias, the hysteresis perpendicular to the field cooling direction does not exhibit any loop shift. The corresponding difference signal of the pinned uncompensated Fe moments vanishes as well.

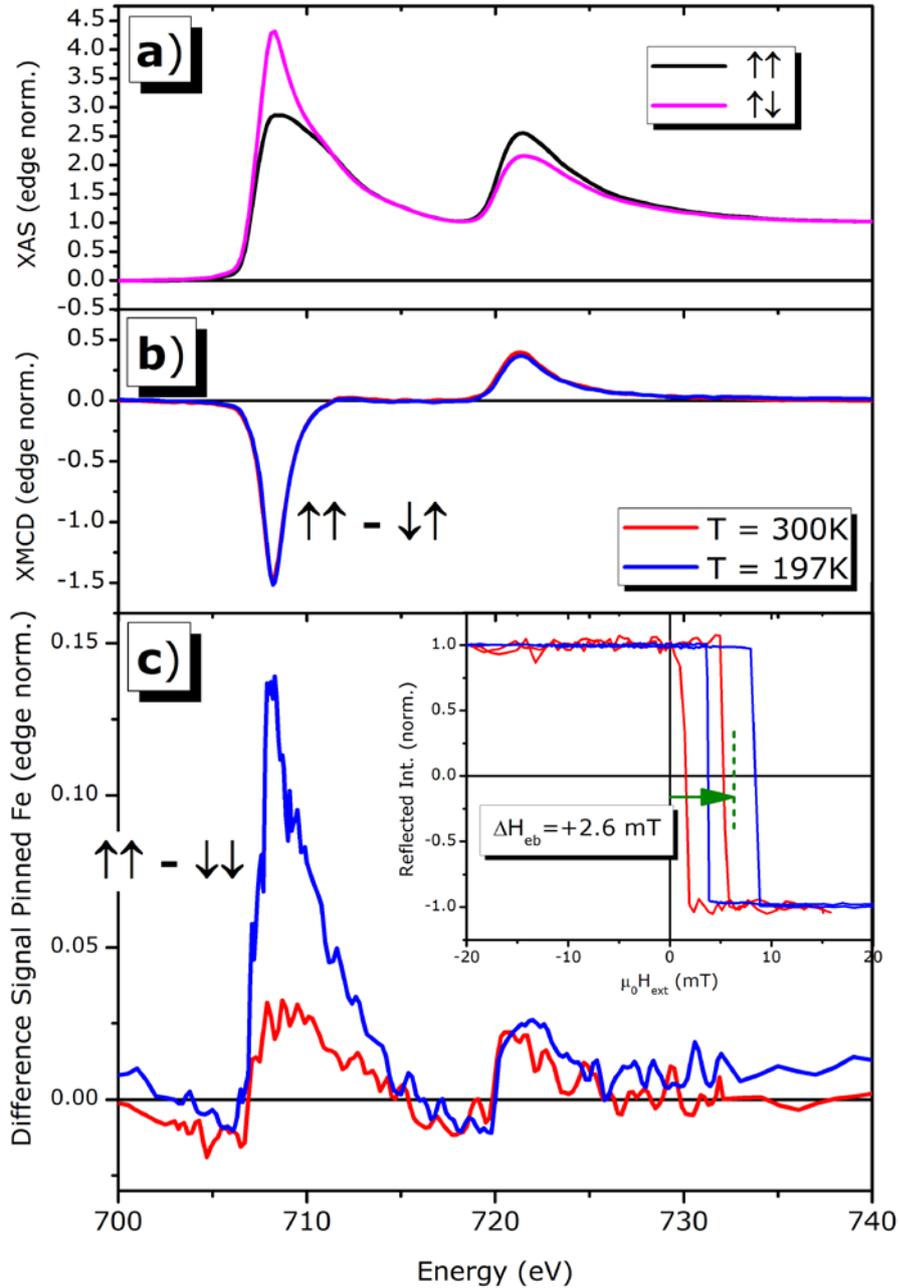

Figure 4: XMCD absorption results. Subfigure a) shows the X-ray absorption spectrum for parallel and antiparallel alignment of X-ray polarization and magnetization measured at the Fe L edges at RT. Figure 4b) shows the corresponding XMCD signal which clearly proves the existence of uncompensated rotatable Fe moments in the sample. A comparison of the RT and 197K XMCD shows no significant differences. The last figure, Figure 4c), shows the signal which is associated to pinned uncompensated Fe at RT and at 197K.